# Magnifying superlens in the visible frequency range.

*I.I. Smolyaninov, Y.J.Hung , and C.C. Davis*

*Department of Electrical and Computer Engineering, University of Maryland, College Park, MD 20742, USA*

Optical microscopy is an invaluable tool for studies of materials and biological entities. With the current progress in nanotechnology and microbiology imaging tools with ever increasing spatial resolution are required. However, the spatial resolution of the conventional microscopy is limited by the diffraction of light waves to a value of the order of 200 nm. Thus, viruses, proteins, DNA molecules and many other samples are impossible to visualize using a regular microscope. The new ways to overcome this limitation may be based on the concept of superlens introduced by J. Pendry [1]. This concept relies on the use of materials which have negative refractive index in the visible frequency range. Even though superlens imaging has been demonstrated in recent experiments [2], this technique is still limited by the fact that magnification of the planar superlens is equal to 1.

In this communication we introduce a new design of the magnifying superlens and demonstrate it in the experiment. Our design has some common features with the recently proposed "optical hyperlens" [3], "metamaterial crystal lens" [4], and the plasmon-assisted microscopy technique [5]. The internal structure of the magnifying superlens is shown in Fig.1(a). It consists of the concentric rings of polymethyl methacrylate (PMMA) deposited on the gold film surface. Due to periodicity of the structure in the radial direction surface plasmon polaritons (SPP) [5] are excited on the lens surface when the lens is illuminated from the bottom with an external laser. The SPP dispersion law is shown in Fig.1(b). In the frequency range marked by the box PMMA has negative refractive index $n_2<0$ as perceived by plasmons (the group velocity is opposite to the phase velocity). The width of the PMMA rings $d_2$ is chosen so that $n_1d_1= - n_2d_2$ , where $d_1$ is the width of the gold/vacuum portions of the interface. While the imaging action of our lens is based on the original superlens idea by J. Pendry, its magnification is based on the fact that all the rays in the superlens tend to propagate in the radial direction when $n_1d_1= - n_2d_2$ (see the inset in Fig.1(b)). This behavior was observed in the experiment upon illumination of the lens with 495 nm laser light (the bottom portion of Fig.1(c)) for which $n_1d_1= - n_2d_2$. The narrow beam visible in the image is produced by repeating self-imaging of the focal point by the alternating layers of positive and negative refractive index materials. On the other hand, if 515 nm light is used the lens becomes uncompensated and the optical field distribution inside the lens reproduces the field distribution in the normal "plasmonic lens" described in ref.[6] (see the top portion of Fig.1(c)).

The magnifying action of the superlens is demonstrated in Fig.1(d). In this experiment three defects have been produced on the superlens surface, as shown in the optical image obtained under white light illumination from the top. When these defects are illuminated from the bottom with 495 nm laser light they produce plasmon beams, which are emitted in the radial direction. The lateral separation between these beams increases by a factor of two as the beams reach the outer rim of the superlens.

To our knowledge, this is the first experimental demonstration of a magnifying superlens in the visible frequency range.

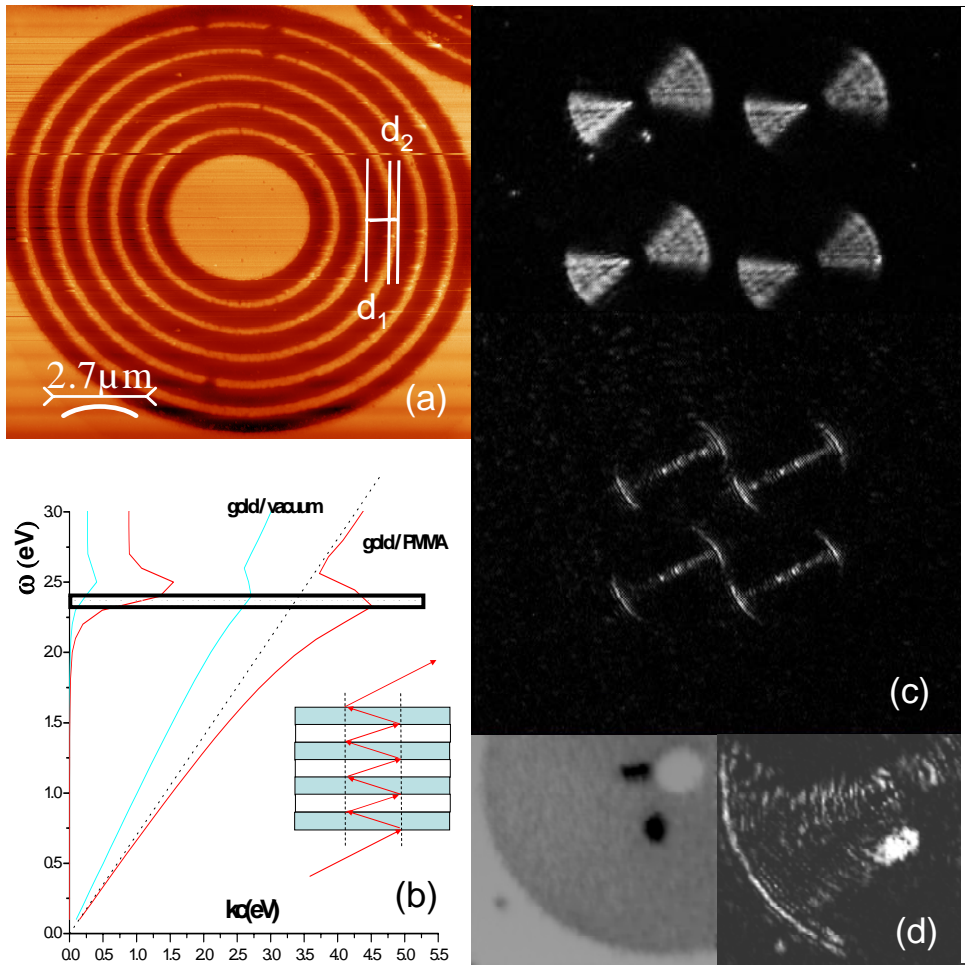

Figure 1: (a) AFM image of the magnifying superlens made of PMMA rings on the gold film surface. (b) Real and imaginary parts of the surface plasmon wavevector at the gold/PMMA and gold/vacuum interfaces as a function of frequency. In the frequency range marked by the box PMMA has negative refractive index as perceived by plasmons. The inset demonstrates that all the rays in the superlens propagate in the radial direction when $n_1 d_1 = -n_2 d_2$. (c) Optical field distribution in the magnifying superlens illuminated by external laser in the frequency ranges where $n_1 d_1 \neq -n_2 d_2$ (top) and $n_1 d_1 = -n_2 d_2$ (bottom). (d) Demonstration of magnification: three defects on the superlens surface emit narrow plasmon beams in the radial directions.